\RequirePackage{fix-cm}
\documentclass[smallextended]{svjour3}       
\smartqed  
\usepackage{graphicx}
\usepackage{caption}
\usepackage{amsmath}
\usepackage{float}
\usepackage{booktabs}
\usepackage{subfig} 
\usepackage{longtable}
\usepackage{multirow}
\usepackage{makecell}
\usepackage{listings} 
\usepackage{soul}
\soulregister\cite7 
\soulregister\citep7 
\soulregister\citet7 
\soulregister\ref7 
\soulregister\pageref7 
\usepackage{color}

\newcommand{\tabincell}[2]{
\begin{tabular}{@{}#1@{}}#2\end{tabular}
}
\begin{document}

\title{AskMe: Joint Individual-level and Community-level Behavior Interaction for Question Recommendation
}


\author{Nuo Li \and 
        Bin Guo \and  
        Yan Liu \and    
        Lina Yao \and   
        Jiaqi Liu \and     
        Zhiwen Yu
}


\institute{Nuo Li \at
              Northwestern Polytechnical University, Xi'an, China \\
              \email{linuo@mail.nwpu.edu.cn}           
           \and
            Bin Guo (corresponding author) \at
            Northwestern Polytechnical University, Xi'an, China \\
            \email{guobin.keio@gmail.com}
           \and
            Yan Liu \at
            Northwestern Polytechnical University, Xi'an, China \\
            \email{253004847@qq.com}
            \and
            Lina Yao \at
            The University of New South Wales, Sydney, Australia \\
            \email{lina.yao@unsw.edu.au}
            \and
            Jiaqi Liu \at
            Northwestern Polytechnical University, Xi'an, China \\
            \email{jqliu@nwpu.edu.cn}
            \and
            Zhiwen Yu \at
            Northwestern Polytechnical University, Xi'an, China \\
            \email{zhiwenyu@nwpu.edu.cn}
}

\date{Received: date / Accepted: date}

\maketitle

\begin{abstract}
Questions in Community Question Answering (CQA) sites are recommended to users, mainly based on users’ interest extracted from questions that users have answered or have asked. However, there is a general phenomenon that users answer fewer questions while pay more attention to follow questions and vote answers. This can impact the performance when recommending questions to users (for obtaining their answers) by using their historical answering behaviors on existing studies. To address the data sparsity issue, we propose AskMe, which aims to leverage the rich, hybrid behavior interactions in CQA to improve the question recommendation performance. On the one hand, we model the rich correlations between the user’s diverse behaviors (e.g., answer, follow, vote) to obtain the individual-level behavior interaction. On the other hand, we model the sophisticated behavioral associations between similar users to obtain the community-level behavior interaction. Finally, we propose the way of element-level fusion to mix these two kinds of interactions together to predict the ranking scores. A dataset collected from Zhihu (1126 users, 219434 questions) is utilized to evaluate the performance of the proposed model, and the experimental results show that our model has gained the best performance compared to baseline methods, especially when the historical answering behaviors data is scarce.
\end{abstract}

\section{Introduction}
\label{intro}
Community Question Answering (CQA) site, such as Quora\footnote{https://quoraconsulting.com/} and Zhihu\footnote{https://www.zhihu.com/}, is a kind of Web service where people can search for information (getting the answers from other people) and share knowledge (answering the questions from other people) online. Compared with traditional information retrieval that only return several related web pages, CQA is more like a high-level information retrieval system, which contains abundant questions and answers that meet users’ need. With the increasing demand of knowledge sharing, the number of questions is continuously increasing and the number of unanswered questions is increasing at a faster rate. For example, Stack Overflow statistics show that the number of questions has a linear increase while the number of unanswered questions has an exponential increase \cite{Ref1}. As a result, it is necessary to solve the problem of question recommendation to guide users to look through the questions they like, make the questions can be answered effectively, and promote the development of the community in a sustainable and harmonious manner.

\begin{figure}[H]
\begin{minipage}{0.5\textwidth} 
\centering
\includegraphics[width=6.5cm,height=4.2cm]{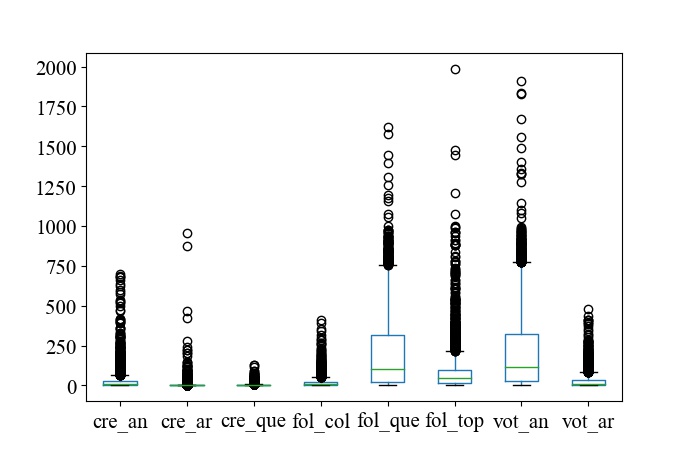}
\caption{\textcolor{red}{\small{Numbers of different behaviors}}}
\label{fig:intro_1}
\end{minipage}
\begin{minipage}{0.5\textwidth} 
\centering
\includegraphics[width=6.5cm,height=4.2cm]{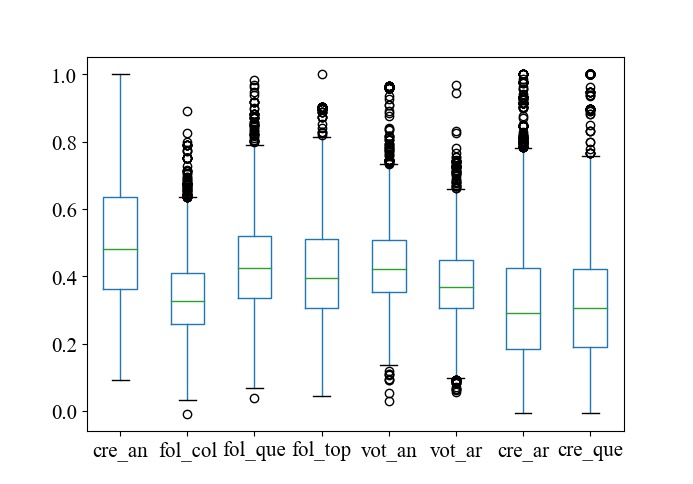}
\caption{\textcolor{red}{\small{Similarities of different behaviors}}}
\label{fig:similarity}
\end{minipage}
\end{figure}

Users only browse or ask questions but not answer questions in CQA. For example, there are eight kinds of user behaviors (creating answer, creating article, creating question, following user, following question, following topic, voting answer, and voting article) on Zhihu platform. And for a certain question, most users follow and vote it while only a few users answer it. We crawl 3996 users of various occupations and educations on the Zhihu platform, and count the number of users’ different behaviors. These statistics are shown in Figure ~\ref{fig:intro_1}. Obviously, the numbers of the following question and voting answer behaviors are more than others. And we can also observe that the two behaviors have higher similarity with the target question in Figure ~\ref{fig:similarity}. Therefore, when users’ data is scarce, these two behaviors can help improve the performance of question recommendation. However, it is obvious that the amount of the answering behavior is the least, with an average of 40 questions per user, while the amount of the following and voting behavior are nearly five times that of answering behavior, reaching an average of 200 questions per user. However, the existing work on question recommendation, including the topic models such as PLSA (Probabilistic Latent Semantic Analysis) and LDA (Latent Dirichlet Allocation) \cite{Ref2,Ref3,Ref4,Ref5,Ref6} and deep leaning method \cite{Ref7} rely on abundant and rich content data to characterize the user's interest better. They are not so effective when the historical answering data is scarce.

Recently, some methods jointly model the auxiliary information associated with users or answers for recommendation to alleviate the data sparseness problem, which mainly apply heterogeneous network containing social relationship. For example, Tu et al. \cite{Ref8} consider social connections to obtain user representation based on that there are strong social connections between users in CQA. Fang et al. \cite{Ref2} and Li et al. \cite{Ref3} propose the heterogeneous network embedding that can collaboratively utilize the rich interaction among the questions, answers and users. These methods mine the relationships among users, questions and answers from the view of the content of questions and answers for recommendation. If a user's social relationships exist, other users connected this user can be regarded as a group, whose interest is similar to this user's interest. However, Such a discrete representation is not enough to show the degree of similarity between users. At the same time, other rich behaviors, such as following and voting behaviors, are rarely considered.

\begin{figure}[htbp]
\centerline{\includegraphics[width=12cm,height=6cm]{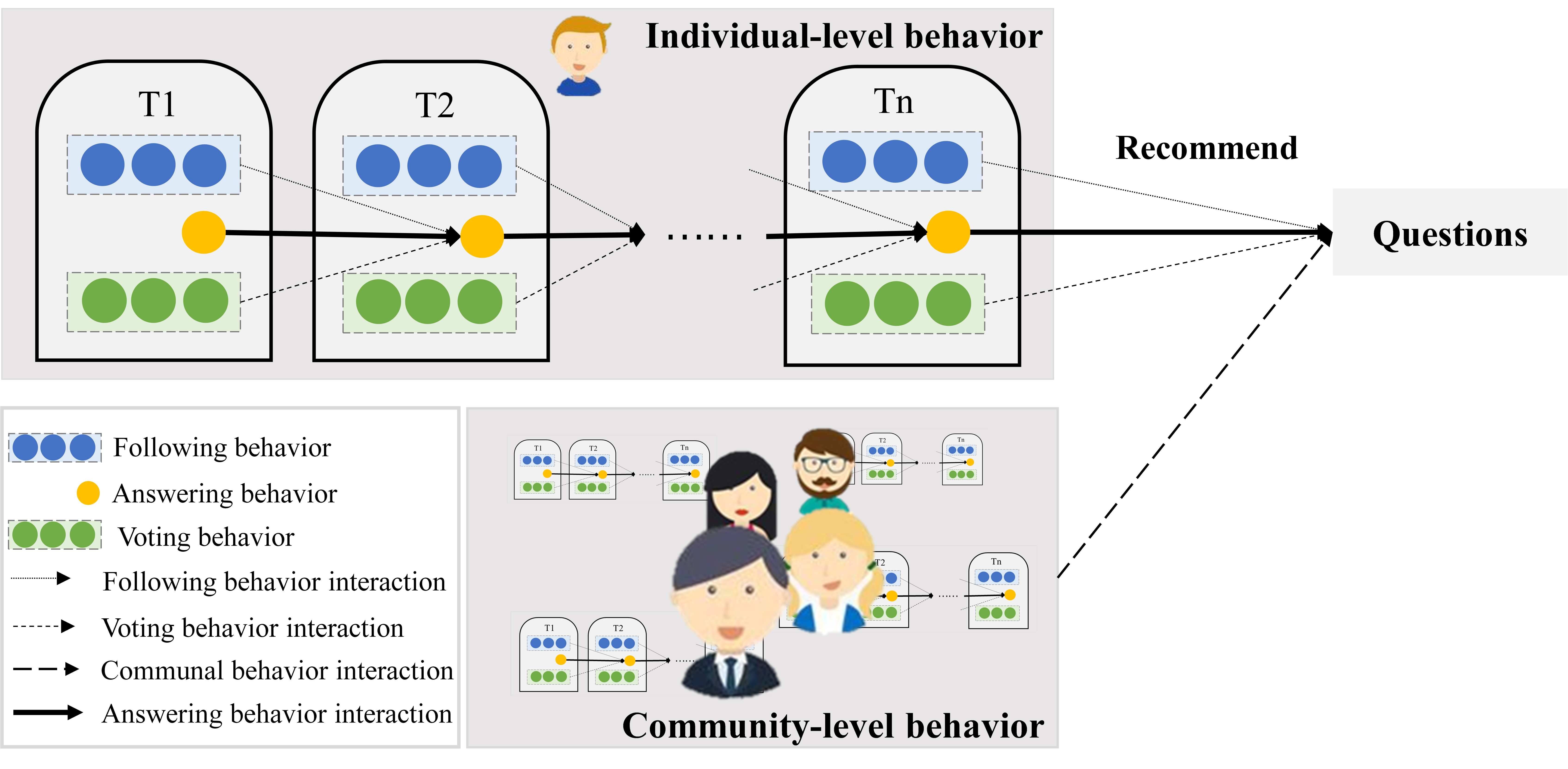}}
\caption{ \textcolor{red}{\small{The case that the individual-level interaction and the community-level interaction can help the user to answer. Here, the individual-level interaction is that the personal interest by modeling the association of following, voting and answering behaviors for a person, and the community-level interaction is the user group's interest based on the personal interest calculated by the individual-level interaction.}}}
\label{fig:two}
\end{figure}

Therefore, we need to tackle two main challenges when facing with the problem of utilizing following and voting behaviors to improve the question recommendation performance as illustrated in Figure ~\ref{fig:two}. The first challenge is \textbf{how to model the individual-level interaction}, i.e., how to deal with the multiple behaviors with huge difference in the amount of data and make them interact with each other. Each kind of the user’s behavior, e.g. following, voting or answering, reflects the user preference to some extent. At present, some studies \cite{Ref9,Ref10,Ref11} model these different behaviors in a same space and obtain the complicated interaction between all behaviors using RNN and other methods to acquire more accurate representation of interest. However, the amounts of user's following and voting behaviors are far more than that of answering behavior as illustrated in Figure ~\ref{fig:intro_1}. Therefore, if we model the three behaviors in the same space, then the interest represented by following and voting behaviors may cover the interest expressed by answering behavior, they cannot promote the possibility of answering, but reduce the probability of answering due to the noise. Consequently, in the case of unbalanced multi-behavior data, how to model the sophisticated associations between behaviors is a great challenge. Another challenge is \textbf{how to model the community-level interaction}, i.e., how to model the impact of behavioral associations of social users. Some studies have demonstrated that considering social relationships improves recommendation performance when the user’s historical behavior data is scarce \cite{Ref8,Ref12}. Generally, these relationships are represented by adjacent matrices, so that the relationship between two persons is either 0 or 1. Such discrete representation indicates that some information is lost to some extent.

For the aforementioned two challenges, we propose the hybrid behavior interaction method, named AskMe, which can fuse the individual-level interaction and community-level interaction for question recommendation in CQA. In particular, in order to deal with the problem of the multiple behaviors with huge difference in the amount of data and make them interact with each other, we propose the individual-level behavior interaction, which can learn the users' main interests from the different behaviors sequences flexibly. In addition, following the spirit of the CF (Collaborative Filtering, CF), we propose the community-level behavior interaction to tackle the problem of the social relationships' influence for question recommendation in CQA. It calculates the group's interests by means of weighted sum. Meanwhile, we regularize the difference between it and the user interest so that the user and the group are more likely to have similar interests. Finally, the individual-level behavior interaction and the community-level behavior interaction are mixed by a concatation and projection operation to predict the click probability simultaneously.

Finally, the main contributions can be summarized as follows:

\begin{itemize}
\item We propose the individual-level interaction method to deal with the problem of multiple behaviors with large difference in the amount of data, which firstly splits the sequence of the following and voting behaviors according to the answering time point and then predicts the results through a multi-view idea.
\item We propose the community-level behavior interaction to handle the problem of the discrete social relationship representation. We calculate the similarity between users based on the users' individual-level interest to describe the continuous relationship representation between users.
\item We propose the AskMe model to combine the individual-level behavior interaction and the community-level behavior interaction based on user's multi behaviors interests to alleviate the problem of data sparsity. 
\item We make experiments on the real-world dataset to evaluate recommendation performance and the experimental results show that our model has gained the best performance, especially when the historical answering behaviors data is scarce.
\end{itemize}

\section{Related Work}
The related work is mainly divided into two parts: question recommendation and behavior modeling.
\subsection{Question Recommendation in CQA Systems}
With CQA system becoming popular in recent years, more and more people study the appealing and challenging problem of question recommendation. And for question recommendation, the goal is to automatically recommend a new question in CQA to the suitable users. Generally, we utilize PLSA, LDA and other topic modeling methods \cite{Ref2,Ref3,Ref4,Ref5,Ref6} to extract the user’s interest distribution from the historical answers or historical questions, and then calculate the relevance between the user’s interest distribution and the target question. Finally, we recommend top k questions to the user according to the relevance scores. For instance, Ni et al. \cite{Ref5} proposed a generative topic-based user interest (TUI) model, which aims to extract the users’ interests by mining the questions they asked and relevant answer providers in the User-Interactive Question Answering (UIQA) systems. Fu et al. \cite{Ref3} considered intimacy between asker and answerer over a topic and proposed a user intimacy model (UIM), which models the relations among the asker, question and answerer, and then the recommendation is obtained by the user’s interest and the intimacy.

With the popular of the deep neural network, representation learning is widely used to extract the features of images or texts  \cite{Ref13}. The word sequence information is also taken into account to learn the semantic representation of the questions in CQA \cite{Ref4,Ref7,Ref8}. Tu et al. \cite{Ref8} employs CNN model to obtain explicit the question textual representation while Zhao et al. \cite{Ref7} and Li et al. \cite{Ref4} utilize LSTM model to get the question features. These methods model the text of the answers or questions, which contains the content and description of the answers or questions. If them are few in CQA, the performance may be poor due to the insufficient extracted interest. 

Facing such a problem of data sparsity, some researchers proposed to apply multi-source data such as social relationships or memory network to improve the recommendation performance \cite{Ref2,Ref7,Ref8,Ref29,Ref30}. Fang et al. \cite{Ref2} proposed the HSNL (CQA via Heterogeneous Social Network Learning, HSNL) model to combine the textual contents with the social relationships to improve the recommendation results. Tu et al. \cite{Ref8} proposed the JIE-NN (Joint Implicit and Explicit Neural Network, JIE-NN) to combine explicit factors and implicit factors based on multiple data sources to deal with the problem of data heterogeneity and sparsity. And Zhao et al. \cite{Ref7} proposed a ranking metric network framework based on users’ relative quality rank to given questions and their social relations to alleviate the data sparsity problem from the viewpoint of learning metric embedding. Sun et al. \cite{Ref29} designed an end-to-end framework that leverages heterogeneous graph and textual information to address the challenge of cold questions. Liu et al. \cite{Ref30} proposed a statical model MAGIC (multi-aspect Gamma-Poisson matrix completion, MAGIC) to automatically generate answer keywords to solve the problem of new questions. Our approach extracts the user’s immediate interest via user’s multiple behaviors and tackles the problem of data sparsity from two aspects: the individual-level behavior interaction and the community-level behavior interaction.

\subsection{Behavior Modeling}
User behavior modeling is a common and challenge problem in the recommendation system. That’s because there are many kinds of user behaviors in the logs, and these behaviors can be applied to analyze the user behavior rules, mine the user preference and predict the user behavior at the moment. Generally, we can divide user modeling into single behavior modeling and multi behavior modeling. Single behavior modeling methods mainly include RNN series (such as DIEN \cite{Ref14}, LSTM and context information \cite{Ref15}, etc.), CNN \cite{Ref16} and Attention mechanism (such as DIN \cite{Ref17}, DSIN \cite{Ref18}, etc.). Multi behavior modeling methods include collective matrix factorization (CMF) \cite{Ref19,Ref20} and modeling into deep semantic spaces together (such as ATRank \cite{Ref11}, NMTR \cite{Ref9}, CSAN \cite{Ref10}, EHCF \cite{Ref31}, MBGCN \cite{Ref32}, etc.). In this part, we focus on multi-behavior modeling, which mainly aims to solve the problem of behavior heterogeneity and data sparsity.

For the data sparsity problem, CMF (Collective Matrix Factorization, CMF) \cite{Ref20} proposed by Singh et al. is the most widely used method, which decomposes the rating matrices for different types of behaviors jointly by sharing the same user latent representation across different behaviors. Cheng et al \cite{Ref19} extended the CMF model and proposed GLFM (Group Latent Factor Model, GLFM) model to integrate multi-behaviors by modeling both the correlation and heterogeneity of them. With the development of deep learning, representation learning is more and more popular. Some studies \cite{Ref10,Ref11,Ref31,Ref32} map all behaviors into a unified low-dimensional space, then the user’s interest can be obtained from the interaction between these behaviors in the low-dimensional space. For example, Zhou et al. \cite{Ref11} proposed an ATRank model based on self-attention mechanism, which maps the heterogeneous behavior sequence into multiple implicit semantic spaces, and selects some important semantics by self-attention to obtain the user’s interest. Huang et al. \cite{Ref10} proposed a CSAN (Contextual Self-Attention Network, CSAN) model, which mapped multi-behaviors into a common latent place, and then the output is fed into specific network to obtain the user’s interest. Chen et al. \cite{Ref31} proposed a transfer-based prediction model EHCF (Efficient Heterogeneous Collaborative Filtering, EHCF) for multiple behaviors with shared embedding matrices and can model fine-grained user-item relations from a semantic perspective. Jin et al. \cite{Ref32} constructed a unified graph to model the multi-behavior data, and proposed MBGCN (Multi-Behavior Graph Convolutional Network, MBGCN) model to capture behaviors’ different influence strength and semantics. These methods make each type of behavior important equally, and not take into account the huge difference in the number of behaviors. 

In conclusion, most methods for multi behaviors recommendation not consider the difference in the amount of various behaviors. At the same time, most of them are multi-task learning, that is to obtain users’ embedding representation and then to learn downstream multiple tasks. While our method cares about the users’ professional knowledge rather than general interests and our approach can select important parts from the abundant following and voting behaviors data, can combine with three data imbalance behaviors. 

\section{Our Proposed Method}
\subsection{Problem Formulation}
We denote $U = \left\{ {u_{1},u_{2},\ldots,u_{|U|}} \right\}$ as the set of users and $I = \left\{ {i_{1},i_{2},\ldots,i_{|I|}} \right\}$ as the set of questions, where $\left| U \right|$, $\left| I \right|$ denotes the number of users and questions, respectively. According to the user’s historical behaviors, we can obtain the set of the user’s historical answers at every moment $ I^{ans} = \left\lbrack  {i_{u1}^{ans},i_{u2}^{ans},~\ldots, i_{u|{ans}_{u}|}^{ans}} \right\rbrack $, where $\left| {ans}_{u} \right|$ denotes the number of historical questions that the user $u$ has answered. For a historical answer $i_{ut}^{ans}$ at any time $t$, we can get sets of the questions that the user $u$ has followed and voted, which are denoted as $I_{t - 1}^{fol} = \left\lbrack {i_{u1}^{fol},i_{u2}^{fol},~\ldots,i_{uN_{1t}}^{fol}} \right\rbrack$ and $I_{t - 1}^{vot} = \left\lbrack {i_{u1}^{vot},i_{u2}^{vot},\ldots,,i_{uN_{2t}}^{vot}} \right\rbrack$, where $N_{1t}$ and $N_{2t}$ denote the number of historical questions that user $u$ has followed and voted before answering the question $i_{ut}^{ans}$. The goal is to predict the next question that the user $u$ may answer according to the questions before the user $u$ has answered at the time $t$, and the user $u$ has followed and voted at the time $t+1$. Table ~\ref{tab:notations} lists the frequently used notations and descriptions in this paper.
\vspace*{-0.5cm}

\begin{table}[htbp]
\caption{\small{Notations and descriptions}}
\label{tab:notations}
\centering
\begin{tabular*}{\hsize}{l|l}
\toprule[1pt] 

\textbf{Notation}       & \textbf{Description}     \\ \hline
$U$    & The set of users \\
$I$   & The set of questions  \\
$I^{k}$   &The set of questions under behavior $k$, $k \in \left\{ follow,~answer,~vote \right\}$  \\
$y_{ui}$ &  The real probability that user $u$ answers question $i$ \\
$\hat{y_{ui}}$  & The prediction probability of the model that user $u$ would answer question $i$   \\
${ei}^{k}$   & The embedding representation of question i under behavior $k$  \\
$\beta_{ui}^{k}$  &  The weight the user $u$ to question $i$ under behavior $k$  \\
$p_{u}^{k}$   & The user embedding representation under behavior $k$     \\
$q_{i}$    &  The representation of the target question $i$      \\
${ei}_{t}^{k}$ & The embedding representation of question $i$ under behavior $k$ at the moment $t$      \\
$\beta_{uit}^{k}$  & The weight the user $u$ to question $i$ under behavior $k$ at the moment $t$      \\
${personal}_{u}$ & The user $u$ embedding in the individual-level behavior interaction      \\
${group}_{u}$ &   The user $u$ embedding in the community-level behavior interaction      \\

\bottomrule[1pt] 
\end{tabular*}
\end{table}

\vspace*{-1cm}

\subsection{Multi-View Model}
\begin{figure}[H]
\centering
\centerline{\includegraphics[width=9cm,height=6cm]{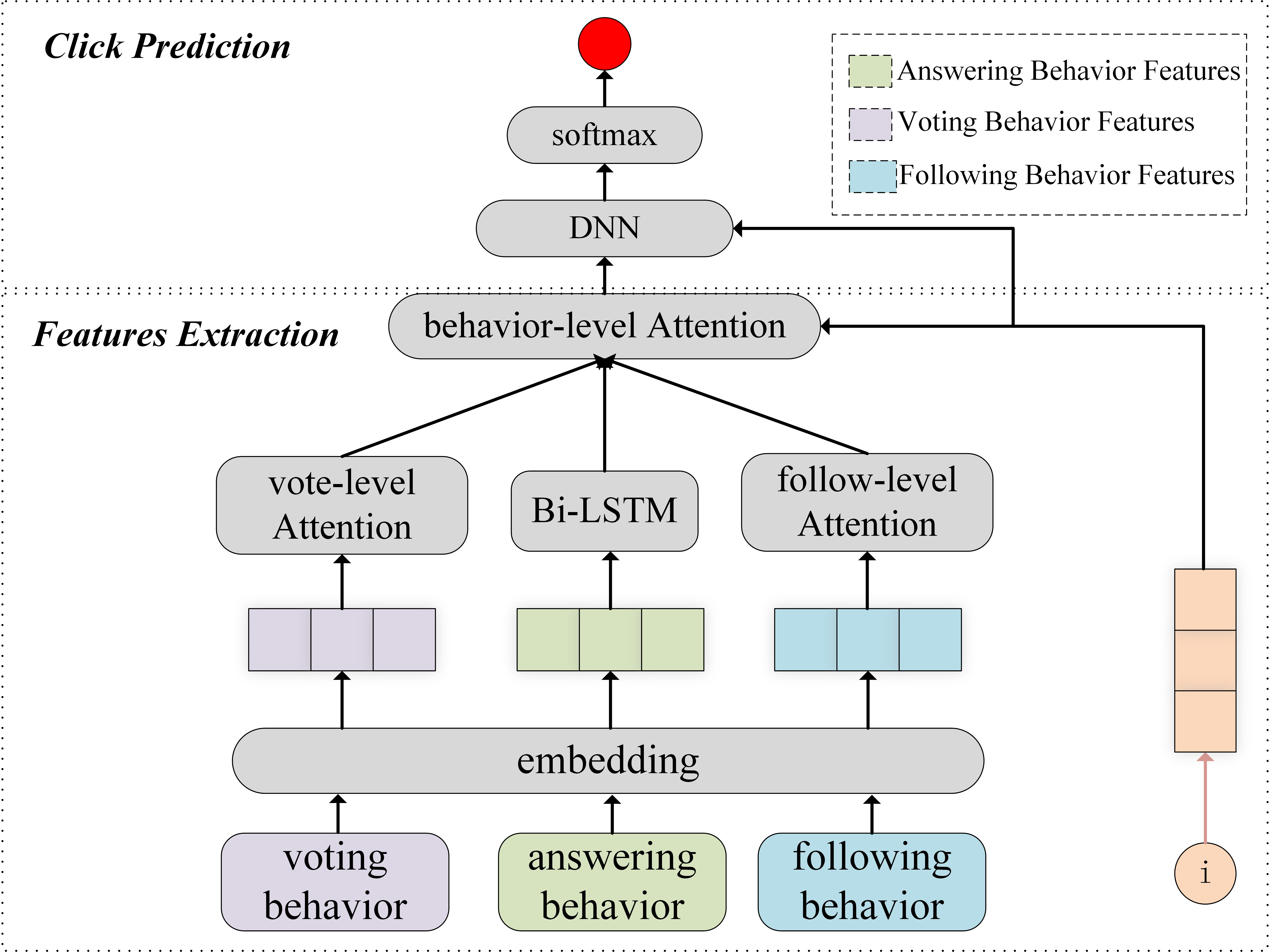}}
\captionsetup{justification=centering}
\caption{ \small{Multi-View Model}}
\label{fig:three}
\end{figure}

There is a phenomenon that data for the answering behavior is scarce while data for the following and voting behavior is rich. Therefore, we improve the recommended accuracy through all three behaviors based on multi-view model \cite{Ref22,Ref23,Ref24}, which is as shown Figure ~\ref{fig:three} . It mainly concludes two parts: features extraction and click prediction. The section of features extraction is to extract the features of these three behaviors, and the click prediction part is to fuse all these features based on a behavior-level attention mechanism to get the final prediction results.

\subsubsection{Features Extraction} 

\paragraph{Answering Behavior Features Extraction}
The historical questions that users have answered indicate the users’ historical interests, which users like, excel at and are keen to answer in the past. Therefore, we model the historical answering questions to express users’ answering preference. As far as we know, LSTM \cite{Ref26} can better capture the long-distance dependence, and can effectively model the historical sequences. However, standard LSTM model can’t capture future information. The user’s answers at any time represent the user’s interest, the user's previous asnwers would have an impact on the current answers,and at the same time, the user's current answers would also produce an effect on the embedding of the previous answers. Therefore, we apply Bi-LSTM to model the historical questions that users have answered. 

\vspace*{-0.2cm}

We denote $I^{ans} = \left\lbrack {i_{u1}^{ans},i_{u2}^{ans},~\ldots,i_{u|{ans}^{u}|}^{ans}} \right\rbrack$ as the set of historical questions that user $u$ has answered, where $\left| {ans}^{u} \right|$ is the sequence length of the historical answering questions. Through an embedding layer , it can be converted into a low-dimension, dense vectors  $\left\lbrack {ei_{u1}^{ans},ei_{u2}^{ans},~\ldots,{ei}_{u{|{ans}^{u}|}}^{ans}} \right\rbrack$. Then the LSTM model is formulated as shown in Eq.(1)-(5):
\begin{gather}
i_{t} = \sigma\left( {W_{x_{i}}{ei}_{ut}^{ans} + W_{h_{i}}h_{t - 1} + b_{i}} \right)   \\
f_{t} = \sigma\left( {W_{x_{f}}{ei}_{ut}^{ans} + W_{h_{f}}h_{t - 1} + b_{f}} \right)    \\
c_{t} = i_{t}{\tanh\left( {W_{x_{c}}{ei}_{ut}^{ans} + W_{h_{c}}h_{t - 1} + b_{c}} \right)} + f_{t}c_{t - 1}  \\
o_{t} = \sigma\left( {W_{x_{o}}{ei}_{ut}^{ans} + W_{h_{o}}h_{t - 1} + b_{o}} \right)    \\
h_{t} = o_{t}{\tanh\left( c_{t} \right)}
\end{gather}

where $i_{t}$, $o_{t}$ and $f_{t}$ denote the input, forget and output gate of the $t$-th object respectively. $W_{x_{i}}$, $W_{h_{i}}$, $W_{x_{f}}$, $W_{h_{f}}$, $W_{x_{c}}$, $W_{h_{c}}$, $W_{x_{o}}$ and $W_{h_{o}}$ are weight matrices. $b_{i}$, $b_{f}$, $b_{c}$ and $b_{o}$ are bias vectors. $c_{t}$ and $h_{t}$ are cell state and hidden state respectively.

We apply Bi-LSTM method, i.e. adding a layer of reverse LSTM to the original forward LSTM network, then we can denote the hidden state of $i$-th as: $h_{i}^{u} = \left\lbrack {\overset{}{h_{i}} \oplus \overset{\leftarrow}{h_{i}}} \right\rbrack$, where $\oplus$ denotes the addition of corresponding elements, $\overset{}{h_{i}}$ denotes the hidden state of $i$-th of the forward LSTM, and $\overset{\leftarrow}{h_{i}}$ denotes the hidden state of $i$-th of the reverse LSTM.

The output of the Bi-LSTM is a sequence $h_{u}^{ans} = \left\lbrack {h_{u1}^{ans},~h_{u2}^{ans}, ~\ldots,h_{u|{ans}_{u}|}^{ans}} \right\rbrack$. We apply $h_{u|{ans}_{u}|}^{ans}$ as the final historical answering questions features, i.e. $p_{u}^{ans} = h_{u|{ans}_{u}|}^{ans}$.

\paragraph{Following Behavior Features Extraction}
The historical questions that users have followed indicate the users’ interest. Due to the scarce answering behavior, we extract users’ current preference from the historical following questions to represent the external factors of users’ answering behavior. We apply a follow-level attention mechanism to capture the relationship between these questions that the user $u$ has followed and the target question, and choose the related questions as the user’ following behavior features.

We denote $I^{fol} = \left\lbrack {i_{u1}^{fol},i_{u2}^{fol},~\ldots,i_{u|{fol}^{u}|}^{fol}} \right\rbrack$ as the set of historical questions that the user $u$ has followed, where $\left| {fol}^{u} \right|$ is the sequence length of the historical following questions. Then it is transformed into a dense vectors $\left\lbrack {{ei}_{u1}^{fol},{ei}_{u2}^{fol},~\ldots,{ei}_{u{|{fol}^{u}|}}^{fol}} \right\rbrack$ via an embedding look-up table $W_{ef} \in R^{V \times D}$ where $V$ and $D$ are the number of all questions and the question embedding dimension respectively. Considering that not all of them affect the final target question, we get important and contextual questions by a follow-level attention method \cite{Ref27} immediately.

The final following features can be denoted as:
\begin{gather}
p_{u}^{fol} = {\sum_{j = 1}^{|{fol}^{u}|}\beta_{uj}^{fol}}ei_{uj}^{fol}   \\
\beta_{uj}^{fol} = \frac{\exp\left( {ei_{uj}^{fol}*q_{i}} \right)}{{\sum_{z = 1}^{|{fol}^{u}|}{ei_{uz}^{fol}}}*q_{i}}
\end{gather}

where $\beta_{uj}^{fol}$ denotes the importance of each question that the user $u$ has followed, and $q_{i}$ is the embedding vector of target question $i$.

\paragraph{Voting Behavior Features Extraction}
The historical questions that users have voted are similar to the historical questions that users have followed, they all indicate the users’ current preference. Therefore, voting behavior features extraction method is akin to following behavior features extraction method. We apply a vote-level attention mechanism to capture the relationship between these questions and the target question, and finally, we get the voting behavior features representation $p_{u}^{vot}$.

\subsubsection{Click Prediction}

The users’ behaviors reflect the users’ preference, and diverse behaviors may lead to various results. For example, the user may answer the questions with same topics compared to the questions that he has followed or voted a kind of question. Therefore, we apply a behavior-level attention mechanism to obtain the weight of each type of behavior and distinguish the role of various behaviors at different times. The representation of the users’ current time can be described as:

\begin{gather}
p_{u} = {\sum\limits_{k = 1}^{3}{\alpha_{k}p_{u}^{k}}}    \\
\alpha_{k} = \frac{exp\left( p_{u}^{k}*q_{i} \right)}{\sum_{j = 1}^{3}{p_{u}^{j}*q_{i}}}
\end{gather}

We denote $\left\lbrack p_{u}^{1},~p_{u}^{2},p_{u}^{3} \right\rbrack$ as $\left\lbrack p_{u}^{ans},p_{u}^{fol},p_{u}^{vot} \right\rbrack$. And the final probability that a user will answer a target question based on their representation is formulated as:
\begin{equation}
\hat{y_{ui}} = \sigma\left( W^{T}\left\lbrack {p_{u};q_{i}} \right\rbrack + b \right),
\end{equation}
where $W$, $b$ and $\sigma$ denote the weight matrix, biases vector and the sigmoid activation function, respectively.

\subsubsection{Objective Function}

To train the model, we use the cross-entropy loss following the probabilistic optimization framework \cite{Ref28} :
\begin{equation}
L = - \frac{1}{N}{\sum_{(x,y) \in D}\left\lbrack ylog\hat{y} + \left( {1 - y} \right)log\left( 1 - \hat{y} \right) \right\rbrack}
\end{equation}
where $D$ represent all the samples, and $y \in \left\{ 0,1 \right\}$ is the label representing whether a user have answered a question.

\subsubsection{Analysis}

The framework is designed to model the user’s three relevant behaviors by the way of the multi-view to capture the association between the candidate item and three behaviors. However, there is no interaction relationship among three behaviors. For example, the following behavior at the moment can help improve the performance of the answering behavior at the next time. Therefore, if we model the interaction at each moment to mine the complicated interaction among three behaviors, the effect of recommendation will be better.

\subsection{The proposed model - AskMe Model}
\begin{figure}[htbp]
\centerline{\includegraphics[width=12cm, height=7cm]{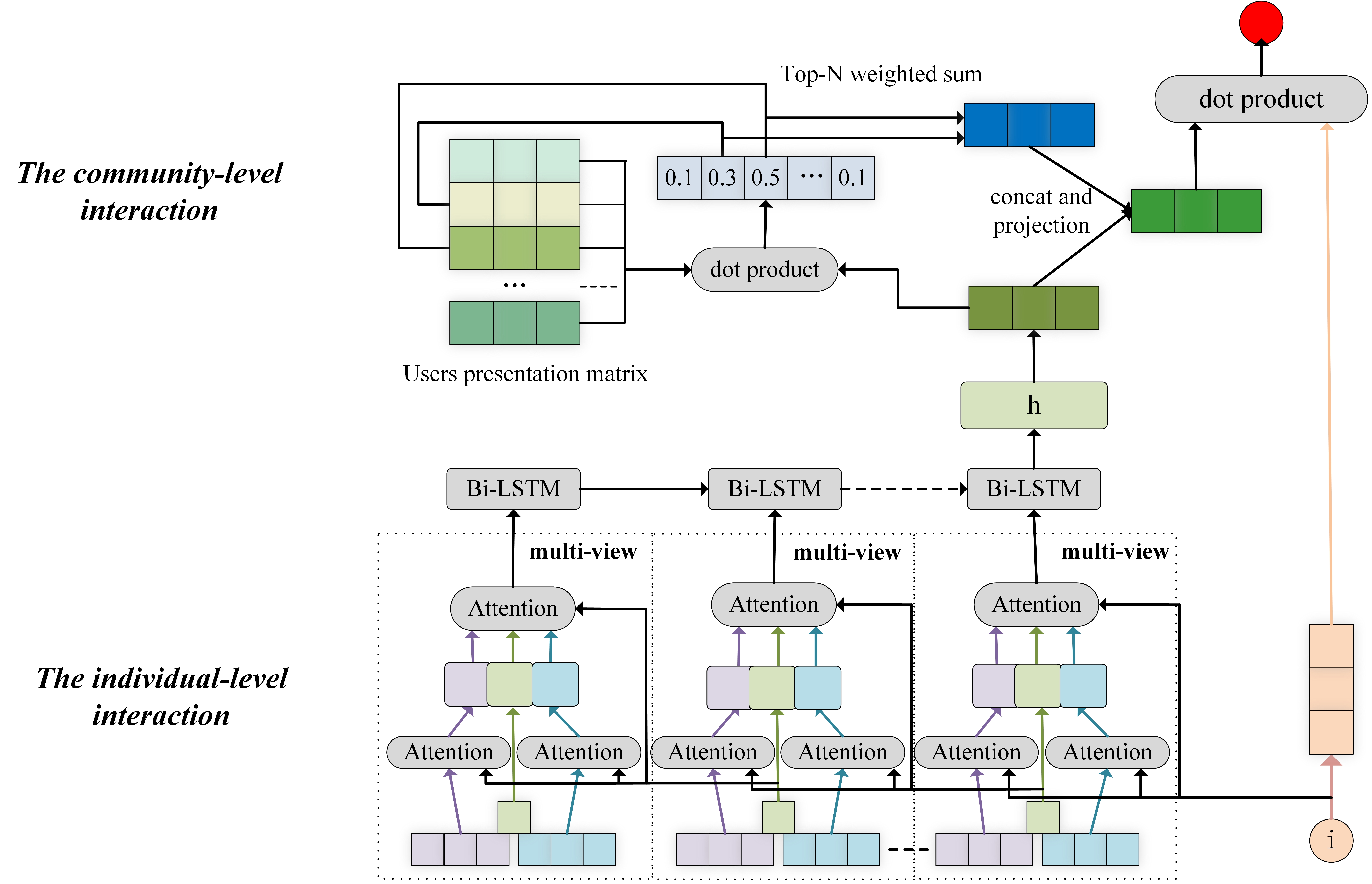}}
\captionsetup{justification=centering}
\caption{ \small{AskMe Model}}
\label{fig:four}
\end{figure}

Considering the interaction at each moment mentioned above, we introduce the thought of the multi-view into each time to get the prediction of the next time. On the other hand, taking other similar people’ s behaviors into account can improve recommendation performance from the perspective of community. In this part, we learn the individual-level behavior interaction and the community-level behavior interaction jointly to tackle the problem of data sparsity. The framework is shown as Figure ~\ref{fig:four}. It mainly concludes two parts: The Individual-level Behavior Interaction and The Community-level Behavior Interaction.

\subsubsection{The Individual-level Behavior Interaction}

For each moment, given the hidden state $h_{ut}$ by LSTM at the time t, the representation vectors $p_{u(t + 1)}^{fol}$ and $p_{u(t + 1)}^{vot}$ of the question that the user $u$ has followed and voted at the time $t+1$, we aggregate them with multi-view model, and then feed it into Bi-LSTM to get the prediction at the time $t+1$. Specifically, the representation vector $p_{u(t + 1)}^{fol}$ can be obtained by:
\begin{gather}
p_{u(t + 1)}^{fol} = {\sum_{j = 1}^{|N_{1(t + 1)}|}\beta_{uj(t + 1)}^{fol}}{ei}_{uj(t + 1)}^{fol} \\
\beta_{uj(t + 1)}^{fol} = \frac{\exp\left( {{ei}_{uj(t + 1)}^{fol}*{ei}_{uj(t + 1)}^{ans}} \right)}{\sum_{z = 1}^{|N_{1(t + 1)}|}{{ei}_{uz(t + 1)}^{fol}*{ei}_{uz(t + 1)}^{ans}}}
\end{gather}
Where $\beta_{uj(t + 1)}^{fol}$ denotes the importance of each question $i$ that user $u$ has followed before the time $t+1$, ${ei}_{uz(t + 1)}^{ans}$ is the embedding vector of answering question at the time $t+1$, and $\left| N_{1(t + 1)} \right|$ is the number of the questions that the user $u$ has followed before the time $t+1$. 

Similarity, the representation vector $p_{u(t + 1)}^{vot}$ is calculated in the same way as the representation vector $p_{u(t + 1)}^{fol}$. Given these three representation vectors: the hidden state $h_{ut}$ at the time $t$, the following vector $p_{u(t + 1)}^{fol}$ at the time $t+1$ and the voting vector $p_{u(t + 1)}^{vot}$ at the time $t+1$, we combine them like a behavior-level attention method, and feed it into Bi-LSTM to get the prediction at the next moment. Finally, we take the last state ${h2}$ as the representation of the user’s interest. And then the final result can be formalized as:
\begin{equation}
{personal}_{u} = ReLU\left( {W_{2}^{T}\left\lbrack {h2;h_{ut};p_{u{\lbrack{t + 1}\rbrack}}^{fol};p_{u{\lbrack{t + 1}\rbrack}}^{vot}} \right\rbrack + b_{2}} \right)
\end{equation}

\subsubsection{The Community-level Behavior Interaction}

Considering the influence of the behaviors of other similar people, we firstly calculate the similarity between the target user and other all users in a dot-product way, which is a widely used method calculating the similarity in a space. If the ${personal}_{u}^{l}$ denotes the representation $l$-th user, then the similarity can be formalized as:
\begin{equation}
\omega_{n} = {personal}_{u} \odot {personal}_{u}^{l}
\end{equation}

Based on that, we select the most similar N users and achieve the representation of similar users’ group by the way of weighted sum. The process of community-level interaction can be formalized as:
\begin{equation}
{group}_{u} = ~{\sum\limits_{l = 1}^{N}{\omega_{l}{\times personal}_{u}^{l}}}
\end{equation}

\subsubsection{Click Prediction}
We combine the two components for deeper and more complicated interaction between the individual-level interaction and the community-level interaction, and make predictions in a dot-product way:
\begin{equation}
\hat{~y_{ui}} = \sigma\left( \left( W^{T}\begin{bmatrix}
{personal}_{u} \\
{group}_{u} \\
\end{bmatrix} + b_{4} \right)\bigodot q_{i} \right)
\end{equation}

\subsubsection{Objective Function}
We still use the cross-entropy loss to train the model. Simultaneously, we regularize the difference between the user representation and the similar user group to make the user and the similar user group have the similar interests. It can be formalized as:
\begin{equation}
L = - \frac{1}{N}{\sum_{{({x,y})} \in D}\left\lbrack {ylog\hat{y} + \left\lbrack {1 - y} \right\rbrack{\log\left\lbrack {1 - \hat{y}} \right\rbrack}} \right\rbrack} + \lambda\left\| {{personal}_{u} - {group}_{u}} \right\|_{2}
\end{equation}

\section{Experiments}
\subsection{Experiment Settings}
\subsubsection{Datasets}
We experimented with a real-world dataset including various kinds of user behaviors collected from Zhihu, which is the largest CQA website in China. The statistics of the dataset are summarized in Table ~\ref{tab:table_data_1}. People of different occupations have different behavioral preferences and diverse interests. Therefore, we also crawled the users of four types occupations on Zhihu website to make our algorithm more convincing and valuable. The data of Computer dataset is the largest, both for numbers of users and questions. Although the data volume of the other three datasets is similar, their sparsity is different (sparsity refers to the number of users multiplied by the number of questions, and then divided by the number of their interactions). In particular, the sparsity of Art dataset is the largest. The statistics of four datasets are listed in Table ~\ref{tab:table_data_2}.

\vspace{-0.1cm}
\begin{table}[htbp]
\caption{\small{Statistics of Zhihu dataset}}
\label{tab:table_data_1}
\centering
\begin{tabular*}{\hsize}{c|cc}
\toprule[1pt] 
\textbf{Type}         & \textbf{Num}    & \textbf{Detail}    \\ \hline
\textbf{users}      & 1,126   & \tabincell{c}{\textbackslash{}} \\
\textbf{questions}   & 219,434 & \tabincell{c}{\textbackslash{}}  \\
\textbf{all interactions} & 746,566 & \tabincell{c}{\textbackslash{}}  \\
\textbf{user properties}  & 8      & \tabincell{c}{nickname\textbackslash{}gender\textbackslash{}avatar\textbackslash{}headline\textbackslash{}education\textbackslash{}business\\employment\textbackslash{}num }   \\
\textbf{user behaviors}   & 8      & \tabincell{c}{create\_answer\textbackslash{}create\_article\textbackslash{}create\_question\textbackslash{}follow\_user\\follow\_question\textbackslash{}follow\_topic\textbackslash{}vote\_answer\textbackslash{}vote\_article }  \\
\bottomrule[1pt] 
\end{tabular*}
\end{table}

\vspace{-0.6cm}
\begin{table}[htbp]
\caption\small{Statistical information of four types datasets}
\label{tab:table_data_2}
\centering
\begin{tabular*}{\hsize}{c|cccc}
\toprule[1pt] 
\textbf{Type}        & \textbf{Art}   & \textbf{Education}   & \textbf{Computer}   & \textbf{Finance}  \\ \hline
\textbf{Number of users}     & 558   & 409   & 1356  & 547  \\
\textbf{Number of questions}   & 153778   & 128904  & 316124   & 150081   \\
\textbf{Number of interactions} & 284606  & 213712   &790734  &278577  \\
\bottomrule[1pt] 
\end{tabular*}
\end{table}

\vspace{-0.2cm}
\subsubsection{Evaluation metric}
In order to evaluate the recommendation performance, we applied two popular metrics, HR (Hit Ratio) and NDCG (Normalized Discounted Cumulative Gain) \cite{Ref28}. 
HR visually indicates whether the question $i$ that user $u$ in the test set would answer in the next time in the recommendation list for the user $u$. 
\begin{equation}
HR@K = ~\frac{hits@k}{N}
\end{equation}
Where $hits@k$ represents the number of the same questions in the test set and the recommended Top-K, and $N$ denotes the number of the questions in the test set.

NDCG shows the ranking quality of the recommendation list.
\begin{equation}
NDCG@K = Z_{n}{\sum\limits_{j = 1}^{n}{\left( {2^{r{(j)}} - 1} \right)/{log\left( {1 + j} \right)}}}
\end{equation}
Where $j$ denotes the position we hope for the goal question in the recommended list and $r\left( j \right)$ denotes the relevance for the position $j$ in the recommended list. In this part, if the recommended question at the position $j$ is in the test set, $r\left( j \right) = 1$, otherwise $r\left( j \right) = 0$. $Z_{n}$ is the normalization coefficient to make the result be in the range of 0-1.

In order to verify the recommendation performance, we adopted the widely used leave-one-out technique \cite{Ref28}, that is, the final answering question is as the test set, and the others as the train set.

\subsubsection{Baselines}
In order to evaluate the recommendation performance of our methods, we compare it with the following baselines:

\paragraph{Question recommendation methods in CQA}
\begin{itemize}
\item \textbf{HSNL} \cite{Ref2}: It encodes not only the contents of question-answer but also the social interaction cues in the community to boost the CQA tasks. Here, we refer to its deepwalk method on user-question interactions to get the latent vectors of the users and questions. Then we employ LSTM to model the relationships among historical behaviors to get the user's preference.
\item \textbf{JIE-NN} \cite{Ref8}: We combine the explicit and implicit information to model user-question interactions. Specifically, we employ CNN from the historical behaviors of a user to get the explicit representation, and get latent user groups representation as the user's implicit representation, then we combine them to get the final results. 
\item \textbf{NeRank} \cite{Ref7}: We apply the deepwalk and word2vec method to learn the representation for the user and question in an interaction view and the embedding of question content from the view of content sequence respectively. There three types embedding as input to fed into convolutional scoring function to get the final results. 
\end{itemize}

\paragraph{Single behavior methods}
\begin{itemize}
\item \textbf{CNN} \cite{Ref16}: We model the users’ historical answering behavior by CNN, and obtain the user historical answering behavior embedding via avg pooling method. We set window sizes from one to five to extract different features and all feature maps to have the same kernel size with 32. 
\item \textbf{DIN} \cite{Ref17}: We use the attention mechanism to obtain users’ different interests.
\item \textbf{DIEN} \cite{Ref14}: We use the GRU method and the attention mechanism to model the evolution of users’ interest and extract users’ preference.
\item \textbf{ATRank} \cite{Ref11}: We use self-attention mechanism to model user answering behavior sequence to get the user embedding. 
\end{itemize}

\paragraph{Multi behavior methods}
\begin{itemize}
\item \textbf{CNN\_M}, \textbf{DIN\_M}, \textbf{DIEN\_M}, \textbf{ATRank\_M}: The above four advanced approaches only model the historical answering behavior. Taking following and voting behaviors into account based on that, we can obtain models CNN\_M, DIN\_M, DIEN\_M and ATRank\_M respectively. Specifically, we sort the three kind of behavioral data according to the time order, and then input the data into CNN, DIN, DIEN and ATRank as a sequence. 
\end{itemize}

\paragraph{our methods}
\begin{itemize}
\item \textbf{AskMe\_M}: The first model we proposed in section 3.2, which computes the interaction among following behaviors, voting behaviors at the last moment and all historical answering behaviors in a multi-view way.
\item \textbf{AskMe\_B}: The model increases the interactions between all behaviors at each moment compared to the AskMe\_M model, so it’s clearly that the individual -level interaction is effective.
\item \textbf{AskMe}: The second model we proposed in section 3.3, which adds the interactions from similar people compared to the AskMe\_B model.
\item \textbf{AskMe\_A}: It denotes that the model contains the historical answered data only, so that we can show the differences between our proposed method and the CNN, DIN, DIEN or ATRank method. 
\item \textbf{AskMe\_P}: It only includes the interactions from other similar people, so we could easily demonstrate that our proposed the community-level interaction is powerful.
\item \textbf{AskMe\_MP}: It contains two parts: AskMe\_M and AskMe\_P.

\end{itemize}

\subsubsection{Parameter Settings}
We implement our and baseline methods in TensorFlow 1.4 and Python 3.6. The experimental parameters are set as follows:
\begin{itemize}
\item \textbf{Initialize approaches}. We utilize Gaussian distribution (mean value is 0 and standard deviation is 0.1) to initialize the parameters for all models
\item \textbf{Network embedding size}. The network embedding size is set to 128.
\item \textbf{Optimizer}. We utilize the Adam, the highly effective optimizer, as the optimizer.
\item \textbf{Batch size}. The batch size is set to 100.
\item \textbf{Learning rate}. We set the initial learning rate to 0.001, which then decays with the decay rate of 0.5 at every epoch.
\item \textbf{Question embedding size}. For each question, we first split the sentence into word sets by Jieba tool, and then calculate the word embedding vector of each word (100 dimensions) by word2vec method. And then we utilize the mean value of word sets in a sentence as the vector representation of this question. In order to make the vector more appropriate, we retrain 28 dimensional vectors, which can be spliced with 100 dimensional vectors into 128 dimensional vectors, as the final representation vector of the question.
\item \textbf{The length of following and voting behavior}. We set the length of following and voting behavior to 5 to save the calculated time and space.
\item \textbf{The number of the similar persons}. We carry out experiments for this number which is set to a range of [3,5,10,20,all] and the performance is the best when the number is 5, so we set it to 5.
\end{itemize}

\subsection{Experiment Results}
\subsubsection{Model Performance Comparison}

\vspace{-0.5cm}
\begin{table}[htbp]
\caption{\small{Evaluation of Top-K question recommendation performance compared to the state-of-the-art approaches under HR with K setting to [10, 20, 30, 40, 50]}}
\label{tab:table_1}
\begin{tabular*}{\hsize}{m{1.38cm}|m{1.6cm}m{1.6cm}m{1.6cm}m{1.6cm}m{1.6cm}}
\toprule[1pt] 

\textbf{Method} & \textbf{Top-10} & \textbf{Top-20} & \textbf{Top-30} & \textbf{Top-40} & \textbf{Top-50} \\ \hline
 HSNL  & 0.339$\pm$0.010         & 0.452$\pm$0.013          & 0.598$\pm$0.011         & 0.723$\pm$0.012          & 0.816$\pm$0.008    \\ 
JIE-NN             & 0.405$\pm$0.013        & 0.558$\pm$0.016         & 0.677$\pm$0.012         & 0.779$\pm$0.014        & 0.835$\pm$0.010         \\ 
NeRank             & 0.333$\pm$0.007        & 0.362$\pm$0.010          & 0.416$\pm$0.010         & 0.501$\pm$0.008         & 0.566$\pm$0.010         \\ \hline
CNN             & 0.360$\pm$0.012          & 0.518$\pm$0.015          & 0.622$\pm$0.013         & 0.702$\pm$0.010        & 0.779$\pm$0.011   \\ 
DIN             & 0.372$\pm$0.009       & 0.505$\pm$0.011         & 0.652$\pm$0.007        & 0.744$\pm$0.011        & 0.809$\pm$0.009        \\ 
DIEN            & 0.429$\pm$0.015         & 0.588$\pm$0.013         & 0.680$\pm$0.014          & 0.769$\pm$0.012         & 0.831$\pm$0.010         \\ 
ATRank          & 0.390$\pm$0.010          & 0.571$\pm$0.011        & 0.675$\pm$0.013        & 0.751$\pm$0.010        & 0.809$\pm$0.007         \\ \hline
CNN\_M          & 0.344$\pm$0.013          & 0.502$\pm$0.010         & 0.604$\pm$0.012          & 0.689$\pm$0.011         & 0.779$\pm$0.009   \\
DIN\_M          & 0.425$\pm$0.010        & 0.585$\pm$0.009         & 0.689$\pm$0.011         & 0.779$\pm$0.010         & 0.840$\pm$0.009          \\  
DIEN\_M         & 0.427$\pm$0.012        & 0.588$\pm$0.015         & 0.685$\pm$0.013         & 0.770$\pm$0.012         & 0.840$\pm$0.011         \\ 
ATRank\_M       & 0.398$\pm$0.009          & 0.576$\pm$0.011         & 0.690$\pm$0.010         & 0.780$\pm$0.008         & 0.838$\pm$0.009         \\ \hline
AskMe\_M        & 0.443$\pm$0.012         & 0.580$\pm$0.012          & 0.710$\pm$0.011          & 0.774$\pm$0.008        & 0.837$\pm$0.011        \\ 
AskMe\_B        & 0.422$\pm$0.010          & 0.590$\pm$0.013         & 0.710$\pm$0.012         & 0.792$\pm$0.013          & 0.850$\pm$0.009         \\  
\textbf{AskMe}  & \textbf{0.469$\pm$0.010}  & \textbf{0.650$\pm$0.011} & \textbf{0.738$\pm$0.009} & \textbf{0.822$\pm$0.007} & \textbf{0.863$\pm$0.010} \\
\bottomrule[1pt] 
\end{tabular*}
\end{table}

We first analyze the top-K performance in our models and all baseline methods. The results are shown in Table ~\ref{tab:table_1} and Table ~\ref{tab:table_2}. It is easy to obverse our final approach AskMe achieves the best performance both under HR and NDCG. At the same time, from the whole point of view, our three methods are superior to the CQA methods, single-behavior methods and the multi-behavior methods. That is because that our methods simultaneously model the interaction of user’s own behaviors and that of other similar users. For CQA methods, HSNL constructs the graph model from the perspective of user-item interaction, and then achieves the initial vectors of users and items in the way of random walk. It's obvious that the more recommended items, the better the recommendation performance. Although the metric for HR is low, the metric for NDCG is high, which shows that the quality of sorting is good, and the model can capture the relative scores of users to the items. As for JIE-NN model, which integrates explicit and implicit information, is better than single-behavior recommendation on HR metric, but worse than multi-behavior and our methods recommendation. There are two possible reasons for the conflict. First, the two metrics cover different aspects: HR estimates the accuracy of prediction, and NDCG calculates the difference between two ranked lists (of a list of candidate items) and emphasizes the location of the items found. Therefore, the changing of them may not coincide under different experiment settings. Second, the candidate items may be similar, so even though the HR value is large, the predicated ranking order of items can be significant difference. This may also lead to conflicts between HR and NDCG results. Meanwhile, it also demonstrates that other behaviors at every moment can effectively improve the recommended accuracy. And for NeRank, an approach based on pairwise loss, its performance on HR is worse than other methods based on pointwise loss, but its performance on NDCG is better. This is also due to the difference between the pointwise loss and pairwise loss. 

\begin{table}[htbp]
\caption{\small{Evaluation of Top-K question recommendation performance compared to the state-of-the-art approaches under NDCG with K setting to [10, 20, 30, 40, 50]}}
\label{tab:table_2}
\begin{tabular*}{\hsize}{m{1.38cm}|m{1.6cm}m{1.6cm}m{1.6cm}m{1.6cm}m{1.6cm}}
\toprule[1pt] 
\textbf{Method}  &\textbf{Top-10}  &\textbf{Top-20} &\textbf{Top-30} &\textbf{Top-40} &\textbf{Top-50} \\ \hline
HSNL             & 0.264$\pm$0.014           & 0.292$\pm$0.013            & 0.323$\pm$0.007          & 0.348$\pm$0.010           & 0.364$\pm$0.009 \\
JIE-NN             & 0.221$\pm$0.011           & 0.265$\pm$0.015         & 0.285$\pm$0.013           & 0.305$\pm$0.009            & 0.315$\pm$0.010           \\
NeRank             & 0.319$\pm$0.012           & 0.326$\pm$0.011          & 0.337$\pm$0.009           & 0.354$\pm$0.012           & 0.366$\pm$0.005           \\ \hline
CNN             & 0.191$\pm$0.011          & 0.231$\pm$0.010           & 0.253$\pm$0.015           & 0.268$\pm$0.009          & 0.282$\pm$0.008 \\
DIN             & 0.214$\pm$0.009        & 0.242$\pm$0.013          & 0.283$\pm$0.009          & 0.288$\pm$0.011           & 0.313$\pm$0.010          \\
DIEN            & 0.249$\pm$0.012            & 0.285$\pm$0.011            & 0.310$\pm$0.007    & 0.325$\pm$0.010            & 0.336$\pm$0.012          \\
ATRank  & 0.211$\pm$0.013           & 0.263$\pm$0.009            & 0.283$\pm$0.010           & 0.300$\pm$0.012           & 0.309$\pm$0.009            \\ \hline
CNN\_M          & 0.173$\pm$0.015          & 0.212$\pm$0.012            & 0.232$\pm$0.014           & 0.249$\pm$0.011           & 0.265$\pm$0.011 \\  
DIN\_M          & 0.250$\pm$0.011           & 0.289$\pm$0.012      & 0.311$\pm$0.008    & 0.327$\pm$0.010  &0.339$\pm$0.007           \\
DIEN\_M         & 0.236$\pm$0.018            & 0.277$\pm$0.009          & 0.297$\pm$0.015          & 0.315$\pm$0.013   &
0.326$\pm$0.010            \\
ATRank\_M       & 0.228$\pm$0.011   & 0.264$\pm$0.010      & 0.280$\pm$0.011            & 0.297$\pm$0.009    & 0.310$\pm$0.007            \\ \hline
AskMe\_M        & 0.257$\pm$0.012   & 0.293$\pm$0.015  & 
0.317$\pm$0.009   & 
0.332$\pm$0.010  &
0.340$\pm$0.010  \\
AskMe\_B        & 0.246$\pm$0.014          & 0.288$\pm$0.010           & 0.318$\pm$0.012           & 0.330$\pm$0.011          & 0.345$\pm$0.009          \\
\textbf{AskMe}  & \textbf{0.280$\pm$0.012}   & \textbf{0.318$\pm$0.010}   & \textbf{0.346$\pm$0.008}   & \textbf{0.357$\pm$0.009}   & \textbf{0.366$\pm$0.007} \\ 
\bottomrule[1pt] 
\end{tabular*}
\end{table}

For single-behavior methods, we observe that DIEN achieves the best performance and CNN obtains the worst performance among the four single-behavior methods. The main reason may be that DIEN models the user’s dynamic interest and extracts the main and proper interest while CNN captures specific sequence patterns, but whether the historical answering behavior contains specific patterns is not sure due to the variability of users' interests.

As for multi-behavior methods, we can see that the trend of results is similar to that of single-behavior methods. However, compared with the performance of single-behavior methods and multi-behavior methods, the result of CNN\_M is worse throughout than CNN. It is because that CNN can’t well mine the main information from the rich and extensive following and voting behaviors, but the attention mechanism can. Therefore, we can find DIN\_M is superior than DIN both on HR and NDCG.

\begin{table}[htbp]
\caption{\small{Evaluation of Top-10 question recommendation performance compared to the state-of-the-art approaches under HR on datasets: Art, Computer, Finance and Education.}}
\label{tab:table_HR}
\begin{tabular*}{\hsize}{m{1.8cm}|m{1.9cm}m{1.9cm}m{1.9cm}m{1.9cm}m{1.9cm}}
\toprule[1pt] 

\textbf{Method} & \textbf{Art} & \textbf{Computer} & \textbf{Finance} & \textbf{Education}  \\ \hline
HSNL  & 0.285$\pm$0.010         & 0.331$\pm$0.013         & 0.293$\pm$0.009          & 0.188$\pm$0.011  \\ 
JIE-NN             & 0.356$\pm$0.020         & 0.398$\pm$0.018         & 0.391$\pm$0.011        & 0.408$\pm$0.008             \\ 
NeRank             & 0.298$\pm$0.012        & 0.337$\pm$0.015         & 0.307$\pm$0.013        & 0.324$\pm$0.010            \\ \hline
CNN             & 0.302$\pm$0.010         & 0.380$\pm$0.013          & 0.300$\pm$0.009          & 0.322$\pm$0.012        \\ 
DIN            & 0.372$\pm$0.009         & 0.396$\pm$0.011          & 0.385$\pm$0.010         & 0.388$\pm$0.008        \\ 
DIEN            & 0.384$\pm$0.011         & 0.410$\pm$0.015          & 0.400$\pm$0.01          & 0.398$\pm$0.010        \\ 
ATRank          & 0.382$\pm$0.006          & 0.401$\pm$0.010          & 0.385$\pm$0.008        & 0.391$\pm$0.007      \\ \hline
CNN\_M          & 0.344$\pm$0.010          & 0.362$\pm$0.015          & 0.305$\pm$0.013         & 0.320$\pm$0.009       \\
DIN\_M          & 0.382$\pm$0.012          & 0.405$\pm$0.009         & 0.386$\pm$0.010         & 0.400$\pm$0.011     \\  
DIEN\_M         & 0.385$\pm$0.011        & 0.418$\pm$0.011          & 0.390$\pm$0.013          & 0.410$\pm$0.015         \\ 
ATRank\_M       & 0.384$\pm$0.011         & 0.413$\pm$0.011          & 0.380$\pm$0.016         & 0.404$\pm$0.012          \\ \hline
AskMe\_M       & 0.384$\pm$0.011        & 0.413$\pm$0.011          & 0.397$\pm$0.016         & 0.404$\pm$0.012   \\
AskMe\_B       & 0.394$\pm$0.017       & 0.427$\pm$0.013          & 0.402$\pm$0.013         & 0.429$\pm$0.015           \\ 
\textbf{AskMe}  & \textbf{0.398$\pm$0.006}  & \textbf{ 0.468$\pm$0.010} & \textbf{0.405$\pm$0.008} & \textbf{0.432$\pm$0.009}     \\

\bottomrule[1pt] 
\end{tabular*}
\end{table}


\begin{table}[htbp]
\caption{\small{Evaluation of Top-10 question recommendation performance compared to the state-of-the-art approaches under NDCG on datasets: Art, Computer, Finance and Education.}}
\label{tab:table_NDCG}
\begin{tabular*}{\hsize}{m{1.8cm}|m{1.9cm}m{1.9cm}m{1.9cm}m{1.9cm}m{1.9cm}}
\toprule[1pt] 
\textbf{Method} & \textbf{Art} & \textbf{Computer} & \textbf{Finance} & \textbf{Education} \\ \hline
HSNL             & 0.231$\pm$0.008            & 0.251$\pm$0.012            & 0.213$\pm$0.015           & 0.228$\pm$0.012     \\
JIE-NN             & 0.201$\pm$0.015            & 0.247$\pm$0.009         & 0.201$\pm$0.007           & 0.212$\pm$0.008               \\
NeRank             & 0.248$\pm$0.010           & 0.267$\pm$0.014          & 0.221$\pm$0.013          & 0.240$\pm$0.008           \\ \hline
CNN        & 0.111$\pm$0.012          & 0.121$\pm$0.015          & 0.173$\pm$0.010           & 0.148$\pm$0.011         \\
DIN             & 0.211$\pm$0.009           & 0.217$\pm$0.008            & 0.221$\pm$0.009           & 0.218$\pm$0.005             \\
DIEN            & 0.220$\pm$0.015            & 0.237$\pm$0.012           & 0.227$\pm$0.011           & 0.234$\pm$0.012          \\
ATRank          & 0.209$\pm$0.007            & 0.222$\pm$0.010            & 0.214$\pm$0.008           & 0.200$\pm$0.009         \\ \hline
CNN\_M          & 0.123$\pm$0.017           & 0.120$\pm$0.020            & 0.160$\pm$0.013            & 0.149$\pm$0.018   \\
DIN\_M          & 0.214$\pm$0.014           & 0.226$\pm$0.009            & 0.210$\pm$0.002          & 0.228$\pm$0.011           \\
DIEN\_M        & 0.226$\pm$0.020            & 0.236$\pm$0.012            & 0.220$\pm$0.012          & 0.223$\pm$0.013               \\
ATRank\_M       & 0.213$\pm$0.013            & 0.224$\pm$0.015            & 0.218$\pm$0.008           & 0.213$\pm$0.009           \\ \hline
AskMe\_M        & 0.226$\pm$0.010   & 0.208$\pm$0.007  & 0.231$\pm$0.008   & 0.233$\pm$0.015   \\
AskMe\_B        & 0.228$\pm$0.017           & 0.236$\pm$0.017            & 0.222$\pm$0.006            & 0.230$\pm$0.010             \\
\textbf{AskMe}  & \textbf{0.244$\pm$0.010}   & \textbf{0.270$\pm$0.012}   & \textbf{0.232$\pm$0.003}   & \textbf{0.252$\pm$0.008}   \\ 
\bottomrule[1pt] 
\end{tabular*}
\end{table}

Furthermore, in order to verify the effectiveness of the model and make the model more convincing, we conducted additional experiments on four types of datasets, and the final results are shown in Table ~\ref{tab:table_HR} and Table ~\ref{tab:table_NDCG} (average and variance after five times). It’s clear that the performance of Computer dataset is the best, followed by Education dataset, and that of Art dataset is the worst. The reason may be that the Computer dataset is the largest, and the sparsity of Art dataset is smallest among the Art dataset, Education dataset and Finance dataset. At the same time, For the dataset with less and sparse data such as Art dataset, AskMe model can get better results, and for the dataset with large amount of data such as Computer dataset, the performance of AskMe model improves more than other models, which indicates that our algorithm has better robustness. 

On the whole, whether for CQA approaches, single-behavior approaches, multi-behavior approaches or our methods, the results of these four types datasets are consistent with the previous results, and our proposed algorithm performance is always the best. and the deviation of our model is least, indicating that our algorithm is most stable. Therefore, our algorithm is more convincing and effective.

\subsubsection{Model Component Analysis}
Compared with some advanced approaches, we can prove that our model is effective in the part \textbf{Model Performance Comparison}. While in this part, we analyze the components of the model to show the effectiveness of our method. The final result is shown as Figure ~\ref{fig:five}. It’s clear that the model with interactions of user’s own behaviors and other similar people’s behaviors has obtained the best performance. Furthermore, we have the following three findings:

\vspace{-1cm}
\begin{figure}[H]
\centering
\subfloat[HR]{
\begin{minipage}{0.47\textwidth} 
\includegraphics[width=\textwidth]{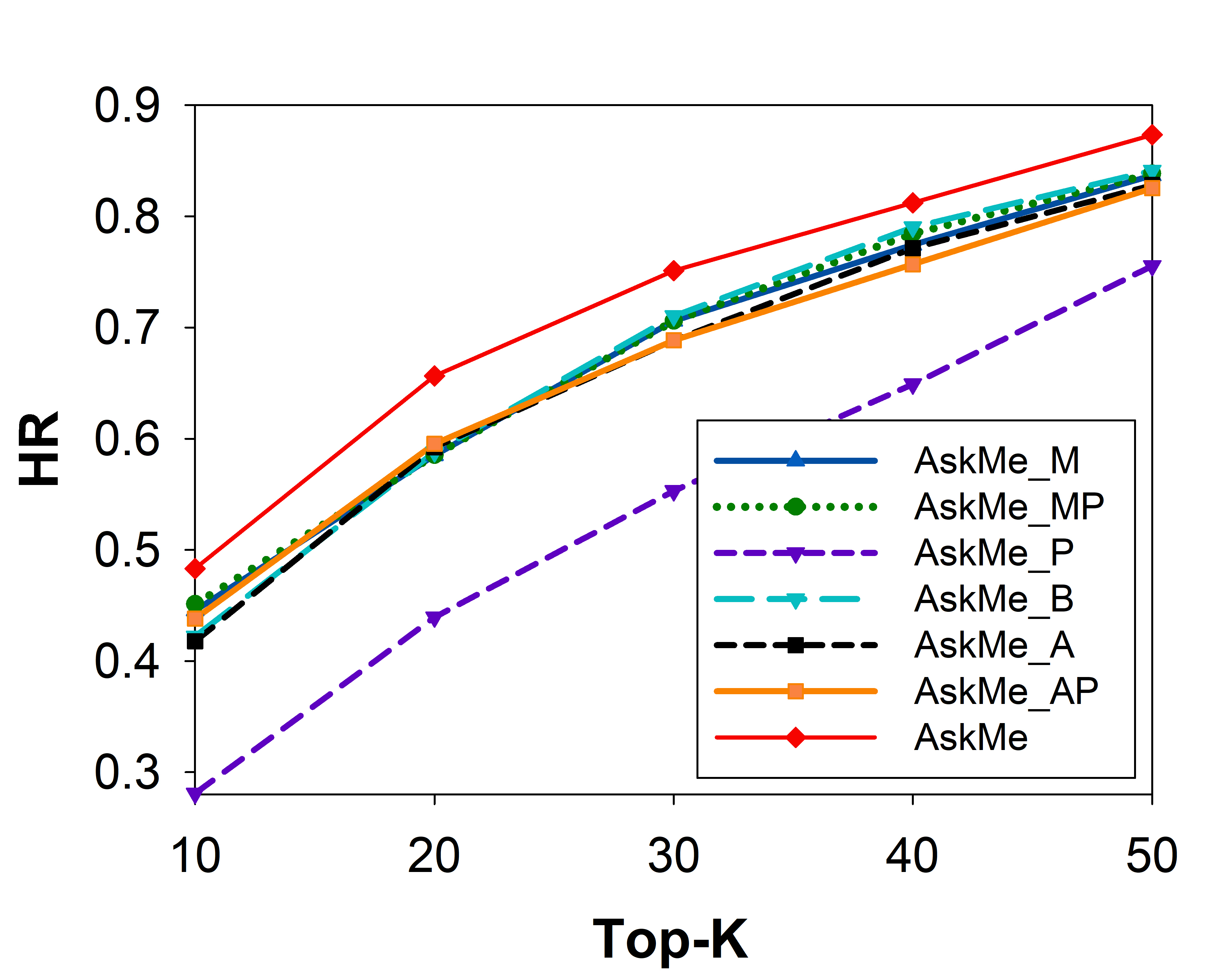} 
\end{minipage}
}
\subfloat[NDCG]{
\begin{minipage}{0.47\textwidth} 
\includegraphics[width=\textwidth]{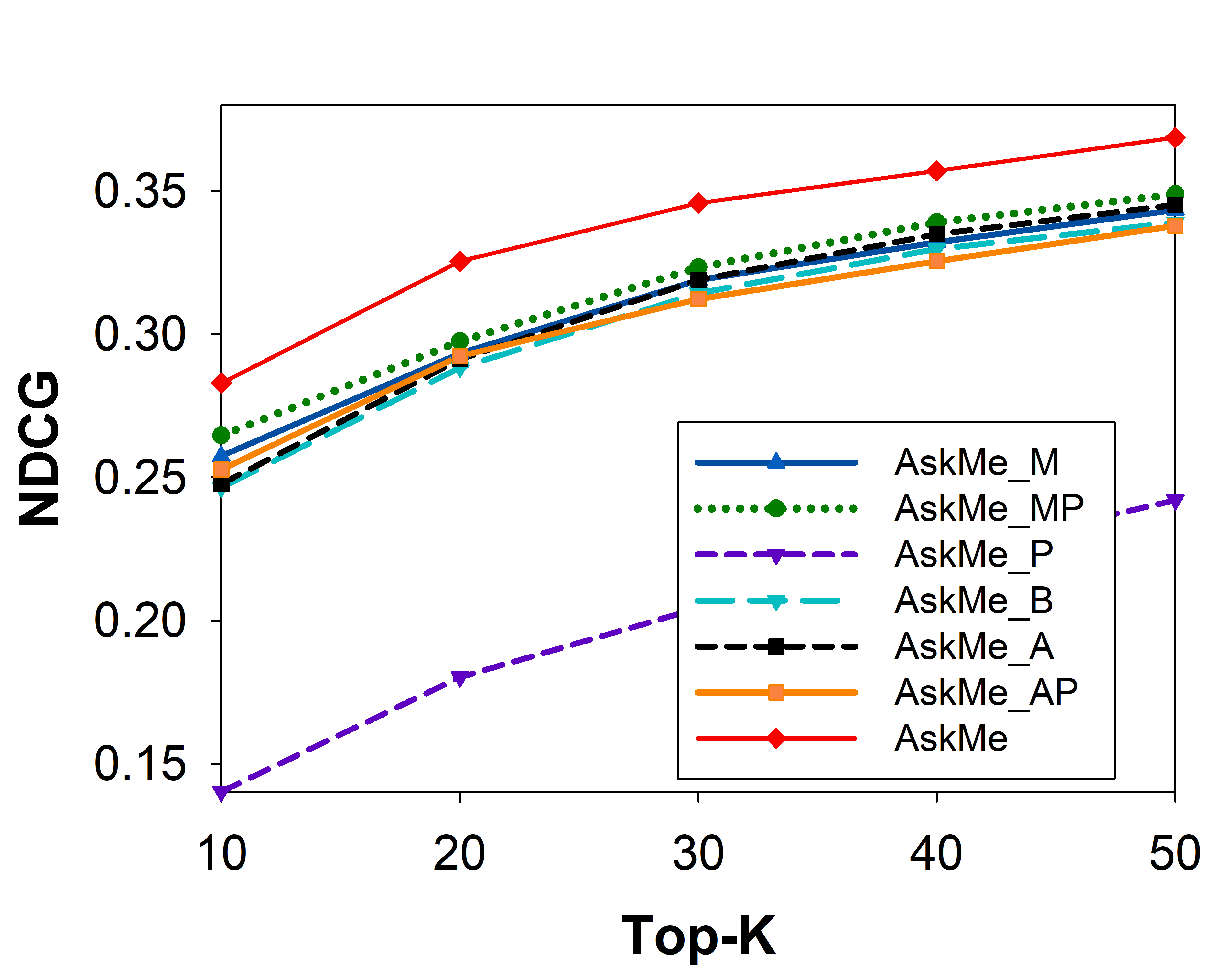} 
\end{minipage}
}
\caption{\textcolor{red}{\small{Evaluation of Top-K question recommendation performance compared to each model component under HR and NDCG}}}
\label{fig:five}
\end{figure}
\vspace{-0.6cm}

\begin{itemize}
\item Observing the results on HR and NDCG for models AskMe\_B, AskMe\_M and AskMe\_A, we can discover that the performance of models AskMe\_B, AskMe\_M both are better than AskMe\_A, which proves the effectiveness of increasing the user’s multiple behaviors.
\item Comparing the performance of model AskMe\_B and AskMe\_M, model AskMe and AskMe\_MP, we can inform that the performance of model AskMe\_B is better than model AskMe\_M on HR, while slightly lower than model AskMe\_M on NDCG, which demonstrates that other behaviors at every moment can effectively improve the recommend accuracy from the perspective of user’s own behaviors. That may be due to the rich and wide following and voting questions. At the same time, we can observe that the performance of model AskMe is better than model AskMe\_MP from alpha to omega, which shows that it’s necessary to increase the component of user’s behavior at each moment.
\item For the interaction component of similar users’ behaviors, we can conclude that the results of model AskMe\_AP, AskMe and AskMe\_MP are always better than model AskMe\_A, AskMe\_B and AskMe\_M from the Figure ~\ref{fig:five}, respectively. Obviously, it’s necessary and very important to increase the interaction component of similar users’ behaviors.
\end{itemize}

\subsubsection{The Length of Answering Behaviors Analysis}
\begin{figure}[H]
\centering
\subfloat[HR]{
\begin{minipage}{0.47\textwidth} 
\includegraphics[width=\textwidth]{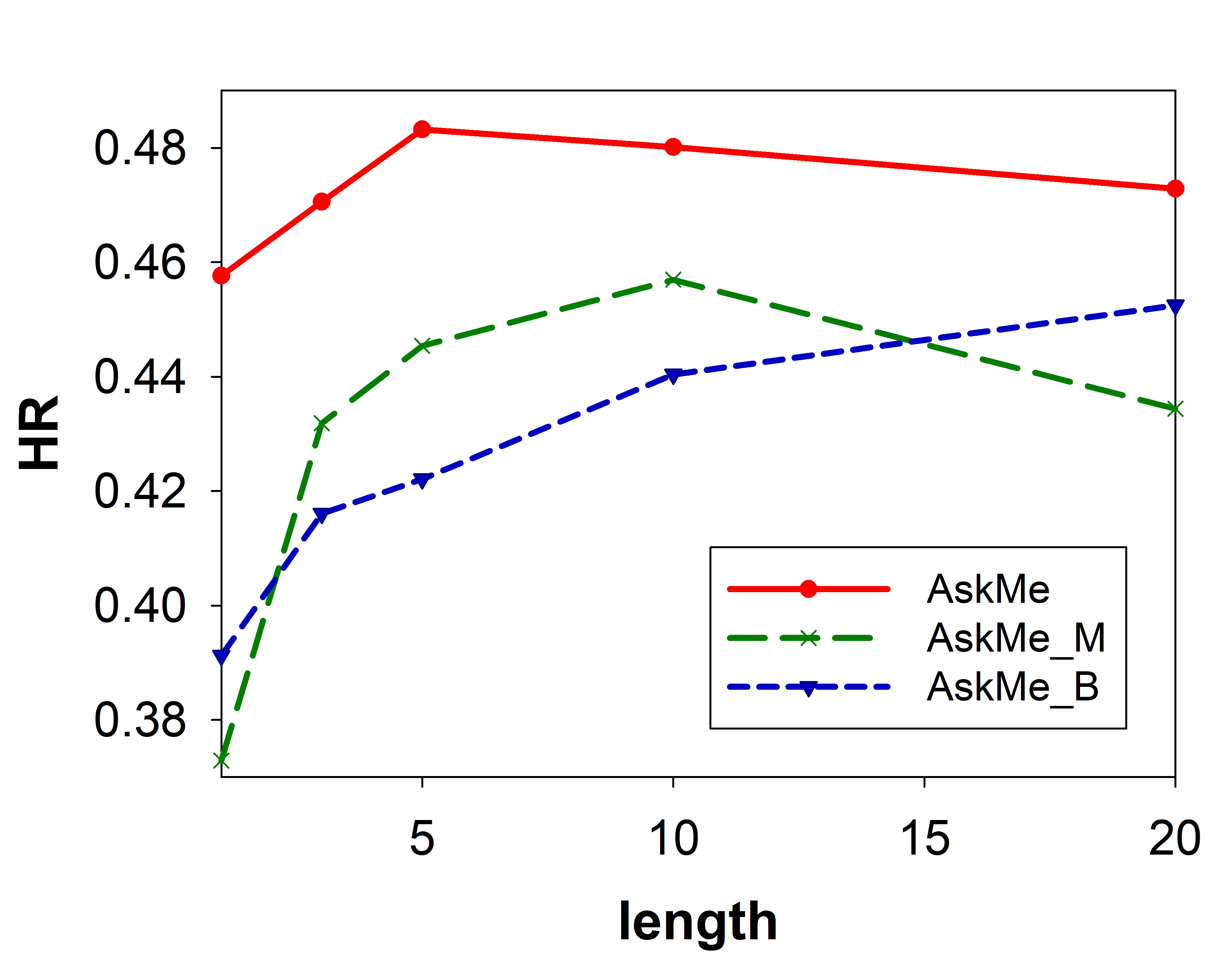} 
\end{minipage}
}
\subfloat[NDCG]{
\begin{minipage}{0.47\textwidth} 
\includegraphics[width=\textwidth]{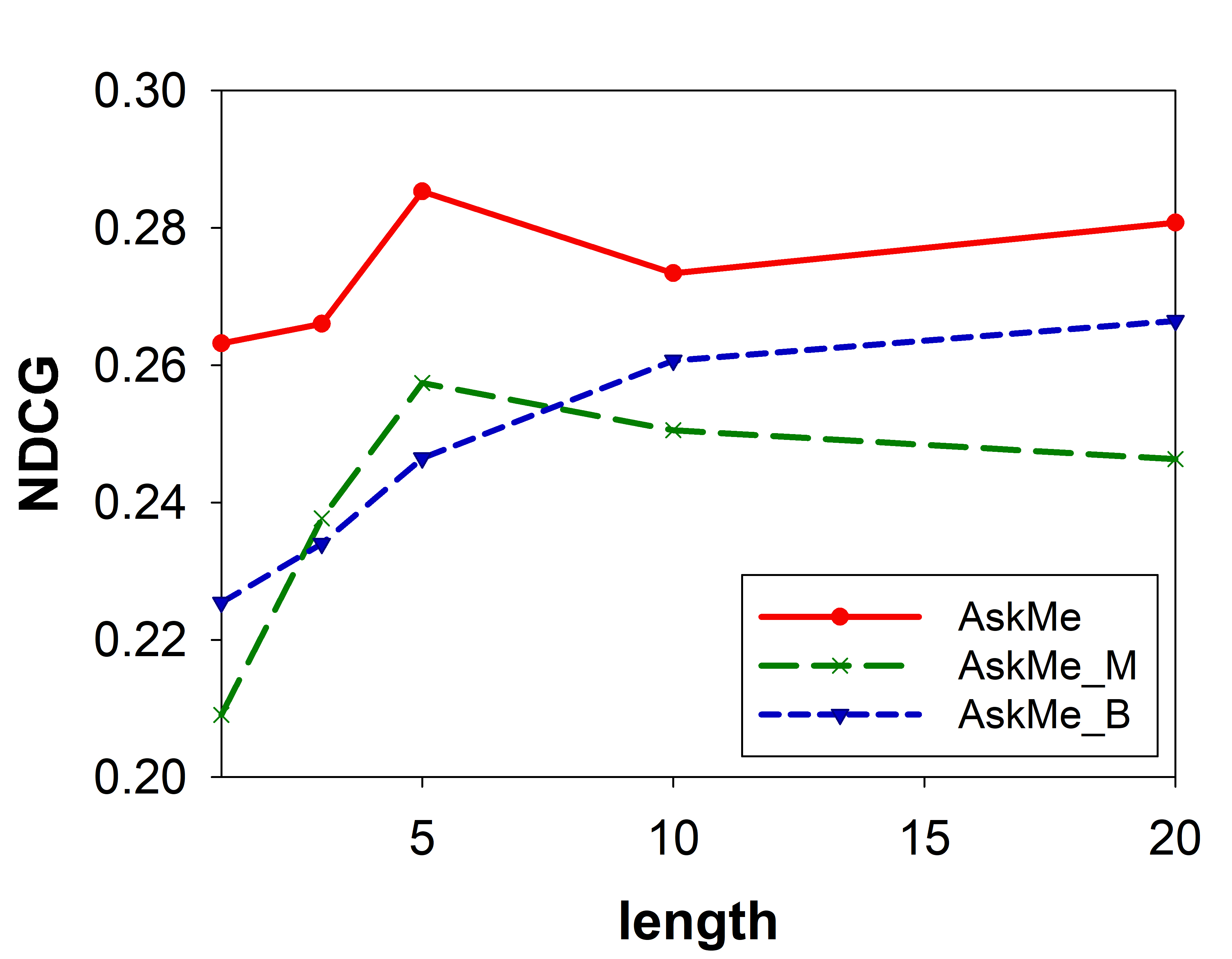} 
\end{minipage}
}
\caption{\textcolor{red}{\small{Evaluation of Top-K question recommendation performance in different lengths [1,3,5,10,20] of historical answering behaviors under HR and NDCG}}}
\label{fig:six}
\end{figure}

The above experimental comparisons are all under the condition that the length of the historical answering behaviors is 5, due to the common phenomenon of scarce answering data in Zhihu website. In this part, we analyze the recommendation performance changes with the length of the historical answering behaviors and the results are showed as Figure ~\ref{fig:six}. It is obvious that our model AskMe always gets the optimal performance no matter what the answering length is. Especially, when the length is less than 5, the performance of this model is far greater than other two methods. At the same time, we can see that the performance of this model is not significantly improved compared with the other two methods when the length is larger than 5. Therefore, there is a conclusion that our model AskMe is very suitable for users with scarce answering data, and the model can obtain good performance when there are 5 historical answering behaviors. 

\vspace{-0.7cm}
\vspace{-0.7cm}

\begin{figure}[htbp]
\begin{minipage}{0.50\textwidth} 
\centering
\includegraphics[width=5.5cm,height=4.0cm]{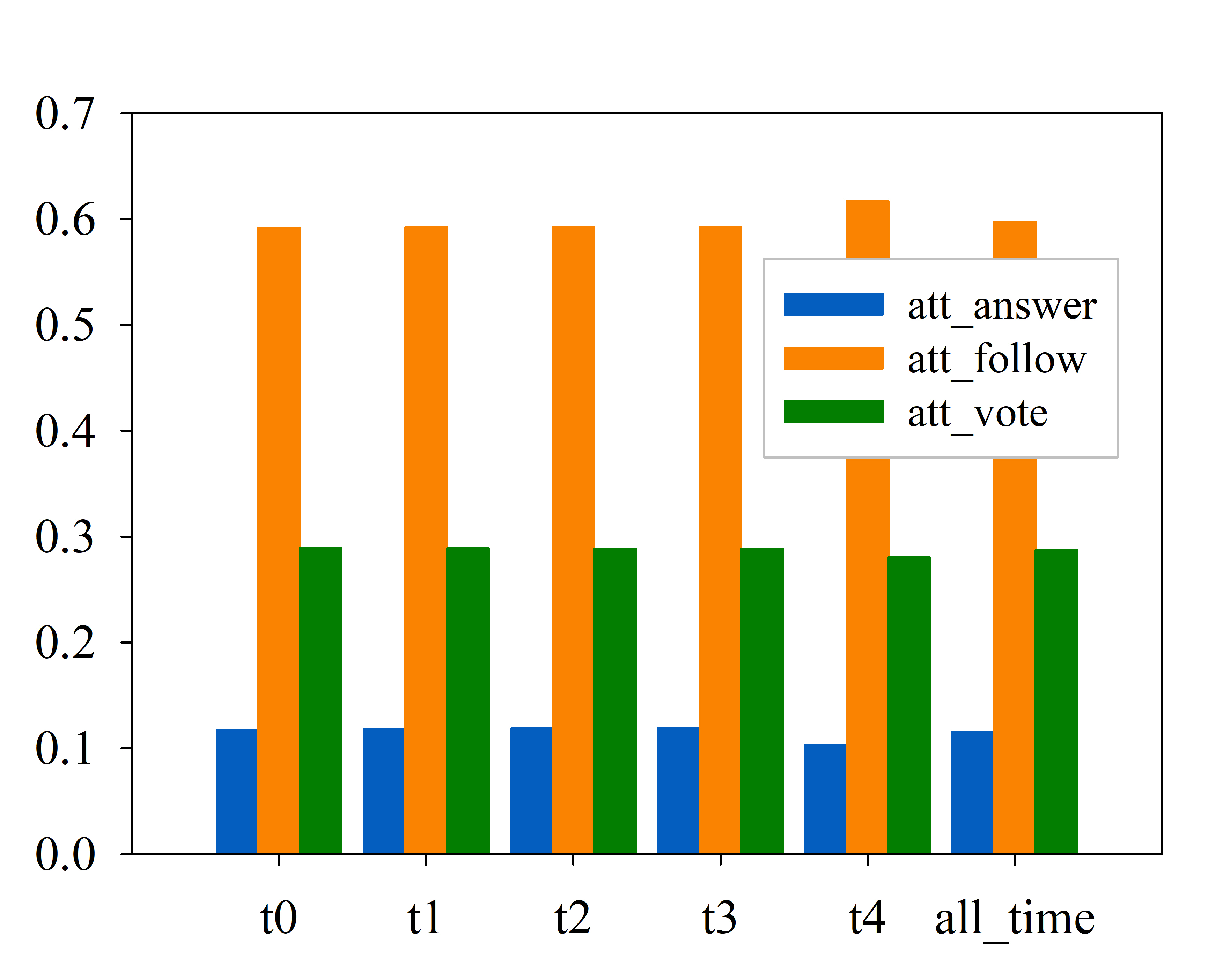}
\caption{\textcolor{red}{\small{The attention scores at each moment}}}
\label{fig:seven}
\end{minipage}
\begin{minipage}{0.50\textwidth} 
\centering
\includegraphics[width=5.5cm,height=4.0cm]{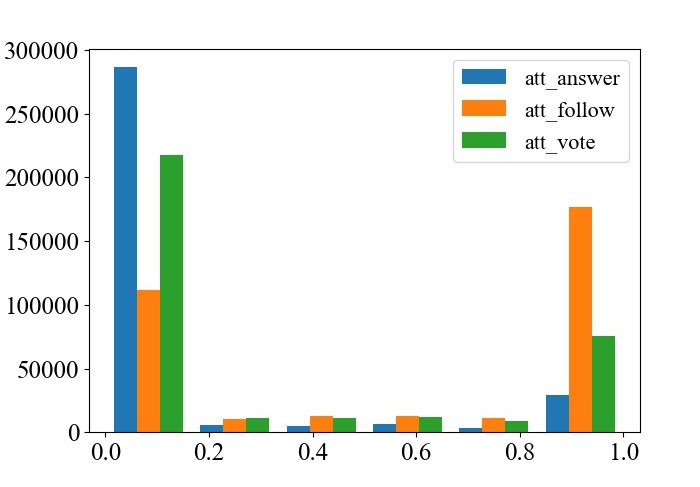}
\caption{\textcolor{red}{\small{The overall attention distributions}}}
\label{fig:eight}
\end{minipage}
\end{figure}

\subsubsection{Behavior-level Attention Analysis}
In order to understand how the various behaviors in the AskMe model interact with each other, we first analyze the mean of attention distribution at each moment, as shown in Figure ~\ref{fig:seven}. From the results, we can observe that the distribution at each moment is extremely similar. And it can be seen that the biggest contribution for the answering behavior at the next moment is the following behaviors no matter what moment. The main reason is probably that the question that the user has followed is what the user likes, then the user is more likely to answer. Therefore, the following behaviors and the voting behaviors at the previous moments will help recommend the questions that the user would answer at the next moment.
As the attention distribution is similar at each moment, we analyze the overall attention distribution, that is, the mean of the attention distribution at all moments, and we can get the result as Figure ~\ref{fig:eight}. From this figure, we can see that most of the attention scores of the user's answering behaviors are mostly in the range of 0-0.2, which indicates that most of the answering behaviors at the previous time do not work for the answering at the next time. While for the attention scores ranged in 0.9-1, it is clear that the number of the following behaviors is largest, which indicates that the users' following behaviors at previous moments have promoted the answering behavior at the next moment. In summary, we can see that the distribution of attention scores for different kinds of behaviors and know that users have various attention scores for diverse behaviors in different situations. All of these just verify the effectiveness of our model.

\subsubsection{Case Study}
\begin{figure}[H]
\centering
\includegraphics[width=12cm,height=4.3cm]{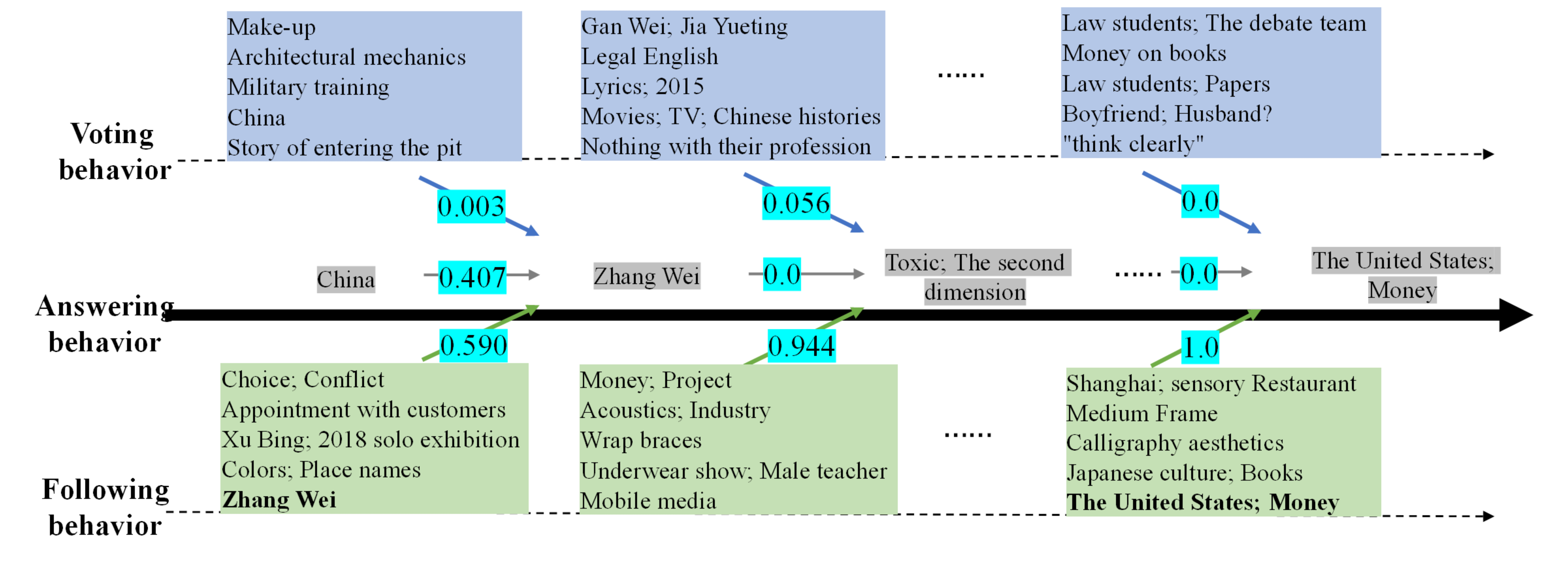}
\caption{\textcolor{red}{\small{The original data and the learned attention scores for the first sample selected}}}
\label{fig:nine}
\end{figure}

The overall attention scores’ distribution of different behaviors is summarized above, and in this section, we randomly select one sample to get the attention score of each behavior at each time, as shown in Figure ~\ref{fig:nine}. It’s easy to see that the attention score of every behavior at each moment is different. In conclusion, our model can select the more important behaviors at each moment to predict the question to be answered at the next moment, and the learned attention scores are consistent with the original data. At the same time, to further illustrate the effectiveness of our method, Table ~\ref{tab:table_case} shows an example of top 3 questions recommended by our methods AskMe, AskMe\_B and another model JIE-NN with best performance in CQA approaches.

\begin{table}[htbp]
\caption{\small{The case study of question recommendation}}
\label{tab:table_case}
\begin{tabular*}{\hsize}{c|p{7cm}}
\toprule[1pt] 
 \multirow{2}{*}{\textbf{{user}}} & userID:midnight-15;   \\
 & occupation:Senior visual designer of Microsoft  \\ \hline
\textbf{\makecell[c]{target question}}   & How to evaluate the wandering earth? \\ \hline
\multirow{5}{*}{\textbf{\makecell[c]{history questions}}} & Q1: How much can people who visit popular art exhibitions understand?  \\
 & Q2: How to evaluate Damien Hirst's work? \\ 
 & Q3: Is art obliged to let others understand your work? \\
 & Q4: Is art obliged to let others understand your work? \\
 & Q5: How to evaluate the wandering earth? \\ \hline
  
\multirow{3}{*}{\textbf{\makecell[c]{Top 3 questions\\ recommended by JIE-NN}}}  & Q1: How to evaluate artistic value for Ni Zan? \\ 
 & Q2: How did you become a music producer/musician? \\ 
 & Q3: what are the top 10 websites for design in China?  \\ \hline
\multirow{3}{*}{\textbf{\makecell[c]{Top 3 questions\\ recommended by AskMe\_B}}}  & Q1: Why are so many science fiction movies related to Mars?   \\ 
 & Q2: who is the artist that affects you the most? \\ 
 & Q3: How to evaluate the wandering earth?   \\ \hline
\multirow{3}{*}{\textbf{\makecell[c]{Top 3 questions\\ recommended by AskMe}}}     & Q1: How to evaluate the wandering earth?  \\ 
 & Q2: Why are so many science fiction movies related to Mars? \\ 
 & Q3: who is the artist that affects you the most?  \\
\bottomrule[1pt] 
\end{tabular*}
\end{table}

\section{Conclusions}

In this work, we propose the AskMe method to solve the problem of scarce answering behavior for the question recommendation in CQA. It can select the important behaviors from rich following behaviors and voting behaviors, and learn the complicated interaction in two aspects: the individual-level behavior interaction and the community-level behavior interaction to predict the question to be answered at the next moment. The extensive experiments are conducted on Zhihu dataset and demonstrate that the complex individual-level and community-level interaction are effective, and the result of the AskMe method is superior to other state-of-the-art approaches on both HR and NDCG, especially in the situation of the scarce historical answering behaviors data.
As for the future work, we will study the multi-behavior recommendation based on multi-behavior data \cite{Ref9}. As learning the main behavior from multiple behaviors improves the recommend accuracy but wastes some resources, such as the time. On the other hand, the time factor is extremely important \cite{Ref21}. For instance, the following behaviors in the first minute may reflect users’ current interests more than that in the first three days, and it is more helpful to predict the question to be answered for users. Therefore, we will add time factor to the model to improve the recommendation accuracy.



\begin{acknowledgements}
This work was partially supported by the National Key R\&D Program of China(2019QY0600), and the National Natural Science Foundation of China (No. 62025205, 62002292, 61725205, 61960206008).
\end{acknowledgements}

%
%



\end{document}